\documentclass[useAMS,usenatbib]{mn2e}
\newcommand{\apj}{ApJ}
\newcommand{\aj}{AJ}
\newcommand{\apjs}{ApJS}
\newcommand{\apjl}{ApJL} \newcommand{\nat}{Nature}
\newcommand{\aap}{A\&A} \newcommand{\mnras}{MNRAS}

\usepackage{latexsym}
\usepackage{mathrsfs}
\usepackage{times} 
\usepackage{graphics}
\usepackage[german,english,american]{babel} 
\def\lesssim{\mathrel{\hbox{\rlap{\hbox{\lower4pt\hbox{$\sim$}}}\hbox{$<$}}}}
\def\gtrsim{\mathrel{\hbox{\rlap{\hbox{\lower4pt\hbox{$\sim$}}}\hbox{$>$}}}}

\title[Jet simulations in dynamic clusters]{The answer is blowing in
the wind: Simulating the interaction of jets with dynamic cluster
atmospheres}

\author[Heinz, Br\"uggen, Young, \& Levesque]{S.~Heinz$^{1}$,
M.~Br\"uggen$^{2}$, A.~Young$^{1}$, E.~Levesque$^{1}$\\$^{1}$Center for
Space Research, Massachusetts Institute of Technology, 77
Massachusetts Avenue, Cambridge, MA 02139; Chandra
Fellow\\$^{2}$International University Bremen, Campus Ring 1, 28759
Bremen, Germany}
\begin{document}
\date{26 June 2006}
\maketitle
\begin{abstract}
We present numerical simulations investigating the interaction of AGN
jets with galaxy clusters, for the first time taking into account the
dynamic nature of the cluster gas and detailed cluster physics.  The
simulations successfully reproduce the observed morphologies of radio
sources in clusters.  We find that cluster inhomogeneities and large
scale flows have significant impact on the morphology of the radio
source and cannot be ignored a-priori when investigating radio source
dynamics.  Morphological comparison suggests that the gas in the
centres of clusters like Virgo and Abell 4059 show significant shear
and/or rotation.  We find that shear and rotation in the intra-cluster
medium move large amounts of cold material back into the path of the
jet, ensuring that subsequent jet outbursts encounter a sufficient
column density of gas to couple with the inner cluster gas, thus
alleviating the problem of evacuated channels discussed in the recent
literature.  The same effects redistribute the excess energy $\Delta
E$ deposited the jet, making the distribution of $\Delta E$ at late
times consistent with being isotropic.
\end{abstract}

\section{Introduction}
\label{sec:intro}
Recent observations show a multitude of physical effects that occur
when active galactic nuclei (AGN) interact with the ambient
intracluster medium (ICM). While these effects are widely believed to
be crucial for the formation of structure in the universe, they are
still poorly understood.

The central galaxy in almost every strong cooling core contains an
active nucleus and a jet--driven radio galaxy. The radio power of
these cooling cores is somewhat correlated with the X-ray luminosity,
although the range of the radio power is much greater than the range
of the X-ray core power. This is supported by a recently discovered
correlation between the Bondi accretion rates and the jet power in
nearby, X-ray luminous elliptical galaxies \citep{allen:06}. These
results show that the AGN at the centres of large elliptical galaxies
feed back enough energy to quench cooling and star formation, thus
providing a possible explanation for the observed cut-off at the
bright end of the galaxy luminosity function \citep{benson:03,
croton:05, bower:06}.  The study of AGN-ICM interactions is of great
interest in a much broader context.  Given the Magorrian relation
\citep{magorrian:98}, clusters hosting a giant cD galaxy should have a
massive black hole in their centre.  The growth of these black holes
is an integral part of feedback--regulated cooling of the ICM.

Radio-loud AGN inflate kpc-sized bubbles of relativistic, underdense
plasma (a.k.a. radio lobes) that displace the hot intra-cluster medium
and thus appear as depressions in the X-ray surface
brightness. High-resolution X-ray observations of cooling flow
clusters with {\em Chandra} have revealed a multitude of X-ray holes
often coincident with patches of radio emission, a compiled list of
which is presented by \citet{birzan:04}.

Numerical simulations of hot, underdense bubbles in clusters of
galaxies have been performed by a number of authors
\citep[e.g.][]{churazov:01, bruggen:02, bruggen:02a, reynolds:01,
ruszkowski:04, dallavecchia:04,omma:04,omma:04b}. Common to all of
these simulations is that they use a spherically symmetric, analytical
profile for the ICM.  \cite{quilis:01} simulated AGN feedback in a
cosmologically evolved cluster but did not model the jets.  Most
recently, \citep{vernaleo:05} performed three-dimensional simulations
of a jet in a hydrostatic, spherically symmetric cluster model. In
their simulation, the jet power was modulated by the mass accretion
rate across the inner boundary. In all of their models, jet heating
failed to prevent the catastrophic cooling of gas at the centre
because the jet preferentially heated gas lying along the jet axis but
failed to heat matter in the equatorial plane. However, in reality the
dynamics of the jet may be strongly affected by the ambient
medium. Bulk velocities may advect and distort the jet and the radio
lobe.  Density inhomogeneities can affect both the jet propagation and
the deposition of entropy in the ICM by the jet. For the purpose of
studying the extent of cluster heating by AGN, the crucial question is
{\em exactly where and how much} energy is deposited by the jet in the
ICM. The primary aim of this work is therefore to study the dynamics
of the jet and the extent of ICM heating subject to the motions and
inhomogeneities of the ICM.

In this letter, we present first results from hydrodynamical
simulations of AGN heating in a realistic galaxy cluster.  The primary
objectives of this work is to study the interaction of the jet with a
dynamic, inhomogeneous ICM.  The letter is organised as follows:
\S\ref{sec:code} describes the technical setup of the simulations,
\S\ref{sec:discussion} presents the results and discusses the
implications for cluster physics, and \S\ref{sec:summary} summarised
and concludes.

\section{Technical setup}
\label{sec:code}
The initial conditions of our simulation are based on a rerun of the
S2 cluster from \cite{springel:01}, whose properties are sufficiently
close to a typical, massive, X-ray bright cluster with a mass of
$M\sim 7\times 10^{14} M_{\odot}$ and a central temperature of 6 keV.
The output of the {\tt GADGET} SPH simulation serves as the initial
conditions for our simulation.  We use the {\tt FLASH} code
\citep{fryxell:00} which is a modular block-structured adaptive mesh
refinement code, parallelised using the Message Passing Interface. It
solves the Riemann problem on a Cartesian grid using the
Piecewise-Parabolic Method. Our simulation includes $7\times 10^5$
dark matter particles. The particles are advanced using a cosmological
variable-timestep leapfrog-method.  Gravity is computed by solving
Poisson's equation with a multigrid method using isolated boundary
conditions.  For the relatively short physical time of the jet
simulation (160 Myrs), radiative cooling and star formation are
neglected, though they were included in the constitutive SPH
simulation.

The cluster shows significant signs of anisotropy as well as some net
rotation of the central cluster gas and contains a dynamically induced
cold-front.  We specifically picked a non-relaxed cluster to
investigate the possible impact of cluster gas dynamics on
jet--cluster interactions, to be compared with hydrostatic, spherical
atmospheres previously investigated \citep[e.g.][]{omma:04}.  The
computational domain is a $2.8 Mpc^3$ box around the cluster's centre
of mass.  The maximum resolution at the grid centre corresponds to a
cell size of $174 pc$, implying 11 levels of refinement.  The total
computing time on our 26 processor cluster amounted to 2 processor
years.  The simulations presented in this letter were performed
assuming an adiabatic equation of state with a uniform adiabatic index
of $\gamma=5/3$ and no radiative cooling.

The jet is injected through a nozzle placed at the centre of the
gravitational potential, coincident with the gas density peak of the
central elliptical galaxy.  The nozzle is modeled as two circular
back-to-back inflow boundaries $2kpc$ or 12 resolution elements in
diameter.  The nozzle faces obey inflow boundary conditions fixed by
the jet's mass-, momentum-, and energy fluxes.  This treatment avoids
the entrainment of cluster gas into the jet which is unavoidable in
simpler schemes where the jet is approximated by injecting mass,
momentum, and energy into a finite volume of the cluster that contains
thermal gas and is part of the active computational grid.  We were
thus able to cleanly separate jet fluid and cluster fluid and to
calculate accurate maps of synchrotron and thermal emission from the
two components, respectively, as well as cleanly measuring the impact
of the interaction on the thermal cluster gas.

VLBI observations of jets \citep[e.g.][]{lobanov:01} and theoretical
arguments about the morphology of radio lobes \citep{scheuer:74} imply
that AGN jets suffer from a range of dynamical instabilities as they
travel through the IGM and their own accumulated exhaust in the lobe.
These instabilities occur on kpc scales and are thus not resolved in
our simulation, leading to measurable deviations from straight, axial
trajectories, which has been dubbed the ``dentist-drill'' effect
\citep{scheuer:74}.  Mimicking it is critical for accurately modeling
injection of momentum and energy by jets.  We included this effect by
imposing a random-walk jitter on the jet axis confined to a 20 degree
half-opening angle.

The jet material is injected equally in opposite directions with
velocity $v_{\rm jet}=3\times 10^{9}\,{\rm cm\,s^{-1}}$ and an
internal Mach number of 32.  The jet power of the simulation presented
in this letter was chosen to be $W_{\rm jet}=10^{46}\,{\rm
ergs\,s^{-1}}$, corresponding to a rather powerful source, comparable
to Cyg A, and in line with the powers now implied by the large scale
moderate shocks found around, e.g., Hercules A \citep{nulsen:05}.  In
order to quantitatively study the jet's effect on the cluster, we ran
a simulation without a jet but otherwise identical parameters for the
same length of time, which we refer to as the control simulation.
Having thus described the numerical setup, we will now proceed to
present and discuss the results.

\section{Results}
\label{sec:discussion}
\subsection{Cluster and Radio Source Morphology}
\begin{figure}
  \resizebox{\columnwidth}{!}{\includegraphics{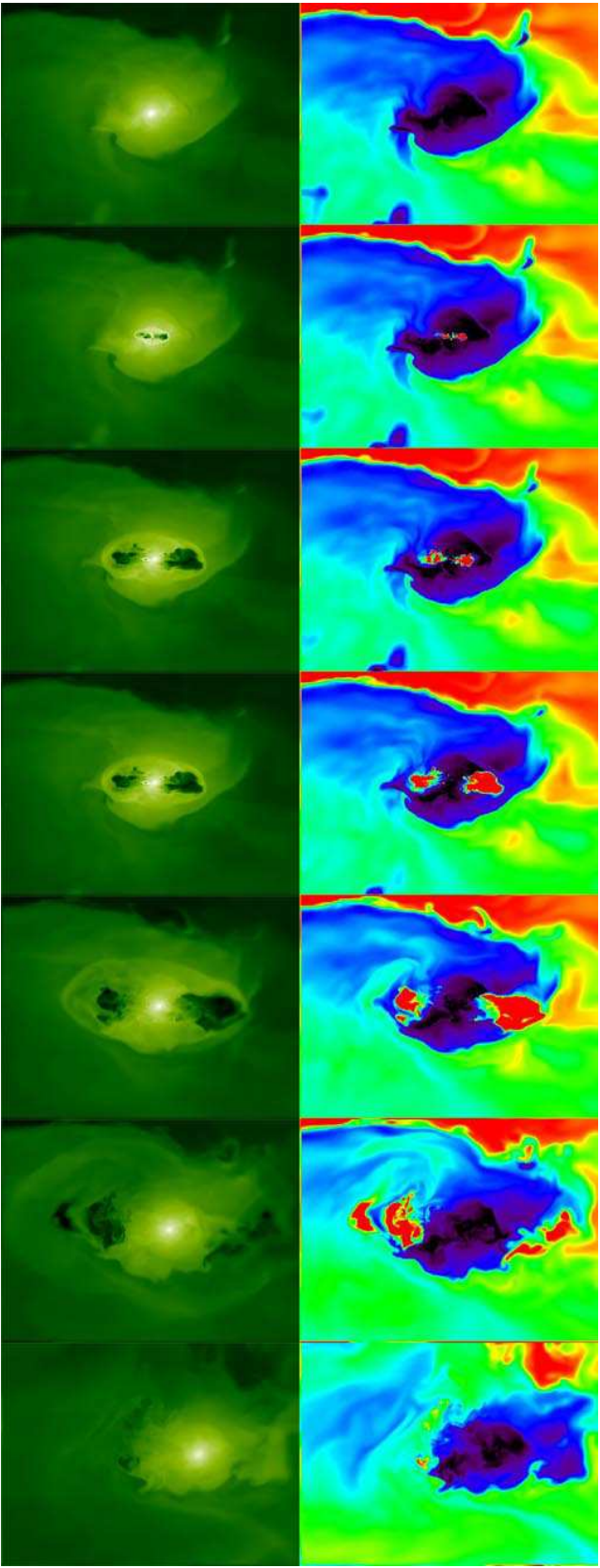}}\
  \caption{Top-to-bottom: Time series of snapshots at 0Myrs, 5Myrs,
    10Myrs, 20Myrs, 40Myrs, 80Myrs, and 160Myrs after jet
    onset. Left-to-right: (a) density cut through cluster centre, (b)
    entropy cut through cluster centre. The images are 450 kpc in
    width and 335 kpc in height.\label{fig:maps1}}
\end{figure}

While the jet is active, the jet-inflated cocoon and the surrounding
swept up shell generally follow the morphological evolution discussed
previously in the literature: The jets inflate two oblong cocoons,
which push thermal gas aside.  The early expansion is supersonic,
shocking the surrounding gas directly, while at later times the
expansion slows down and eventually becomes sub-sonic.
Fig.~\ref{fig:maps1} shows a time series of density and entropy cuts
as well as simulated X-ray maps (using the Chandra ACIS-I response and
assuming an APEC thermal model) and low-frequency radio maps (assuming
equipartition and neglecting radiative cooling).  The aspect ratio of
the lobes is roughly 3 and decreases with time, indicating the lateral
spreading and shear of the radio plasma.

\begin{figure}
  \resizebox{\columnwidth}{!}{\includegraphics{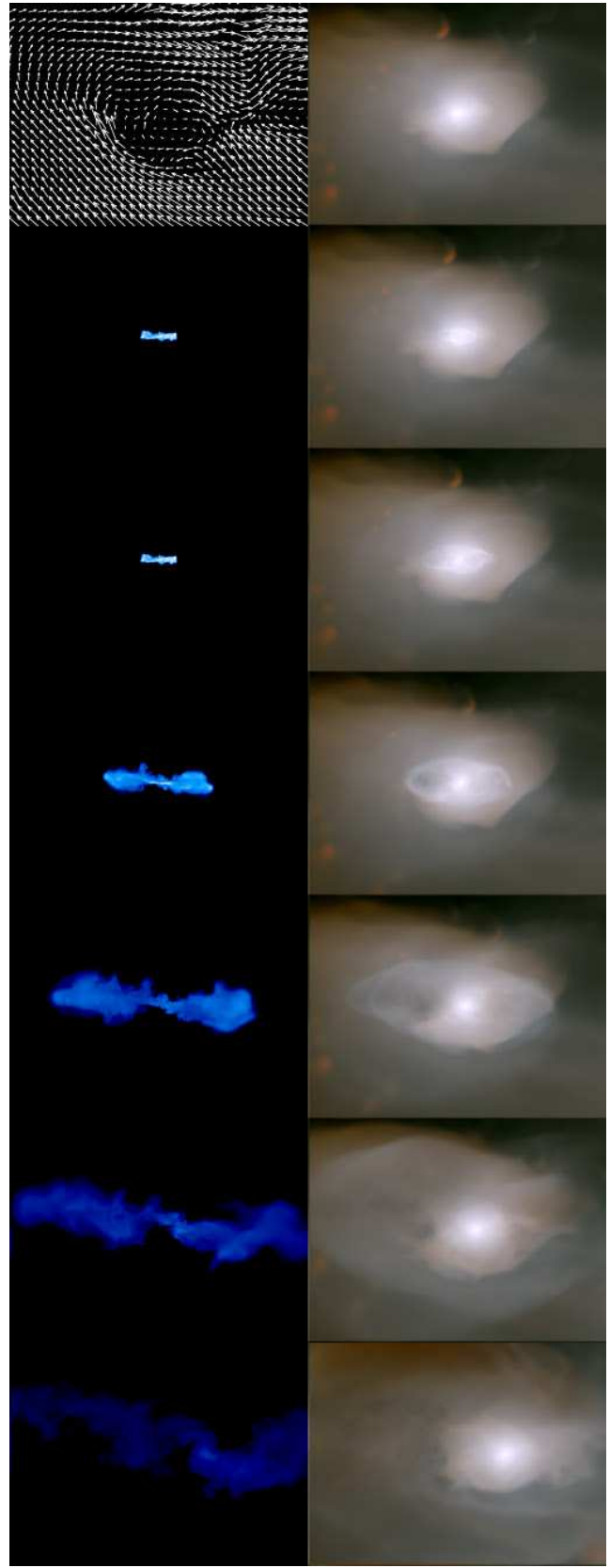}}
  \caption{Time series of snapshots (0Myrs, 5Myrs, 10Myrs, 20Myrs,
    40Myrs, 80Myrs, and 160Myrs after jet). Left-to-right: (a) Top
    panel: velocity map through cluster center; below: low-frequency
    radio synchrotron map, (b) X-ray maps (red: 0.3-2\,keV, green:
    2-5\,keV, blue:\,5-10 keV).\label{fig:maps2}}
\end{figure}

During the active source evolution, while the jet is on, the source
closely resembles the best studied example of a source of comparable
power: Cyg A.  The fact that aspect ratio and X-ray and radio
morphology of the simulated source so closely resemble Cyg A indicates
that we properly modeled the effect of the dentist drill effect (which
can, in fact, be directly observed in Cyg A, where the jet axis is
clearly mis-aligned with the radio and X-ray hot spots).  We verified
this by running a test case with half the opening angle of the jitter
cone ($10^{\circ}$ instead of $20^{\circ}$) which produced too narrow
a cocoon (with an aspect ratio of around 7).  

The X-ray maps show a clear increase in the surface brightness in the
outgoing shock/compression wave and a wispy, turbulent wake behind the
wave (Fig.~\ref{fig:maps2}).  This is best demonstrated by un-sharp
masking of the images.  An unsharp masked movie of the X-ray images
can be viewed at {\tt http://space.mit.edu/$\sim$heinzs/jetpower/},
showing a network of roughly spherical outgoing sound waves that are
excited behind the major initial shock wave, consistent with the
complex network of sound waves seen in Perseus \citep{fabian:03}.
They also show bud-like bubbles appear at the edges of the main shell
when the jet axis moves due to ``denstist-drill'', reminiscent of the
budding bubbles in M87 \citep{forman:05}.

Two distinctions to previously published results are
important to point out regarding the morphological evolution of the
source:

First, it is clear from the series of snapshots shown in
Fig.~\ref{fig:maps1} that the non-sphericity of the atmosphere induces
significant asymmetries in the two sides of the cocoon.  Given that
the difference is so strong, it is clear that the impact on the
morphology as a whole must be significant.  This justifies our initial
motivation for this study: In order to investigate the evolution of
radio sources and their impact on galaxy cluster atmospheres, one
cannot neglect the dynamic nature of the atmospheres prior to jet
injection.

Second, the dynamic nature of the cluster atmosphere significantly
alters the late stage evolution, after the jet switches off (between
the fourth and the fifth frame at $3.3\times 10^{6}\,{\rm yrs}$) and
the source evolution becomes sub-sonic.  During the sub-sonic stage,
the impact of the pressure and density of the surrounding gas become
more and more important.  In particular, the rotation that is present
in the cluster (see Fig.~\ref{fig:maps2}) has clearly sheared the
plasma away from the jet axis.  We will quantify these results below.
It is already clear, however, that this effect can solve the dilemma
of launching multiple jet episodes into the same cluster: Any
evacuated channel created by previous outbursts will have been
buffeted around sufficiently to move enough cluster gas back into the
path of the new jet to provide a sufficient cross section for
interaction and to couple the jet to the inner cluster gas.

The spiral-like morphology of the radio source at late stages is very
reminiscent of the large scale ($>30\,{\rm kpc}$) morphology of M87
and Abell 4059 \citep{heinz:02}, where the large scale radio bubbles
are clearly mis-aligned with the inner jet.  This leads us to suggest
that there might be significant rotation or shear present in the gas
of these clusters.

Before proceeding to a more quantitative analysis, it is worth
pointing out that the episode of jet activity {\em does not} disrupt
the cluster or blow the central region apart, despite the fact that we
are simulating a powerful source (certainly at the upper end of the
power range expected from central cluster radio sources).  Contrary to
what might be expected naively, the expansion shock weakens and does
not super-heat the central atmosphere beyond convective stability.
This does not mean, however, that the jet does not induce significant
turbulent motion in the central cluster.

\subsection{Quantitative results}
\begin{figure}
  \resizebox{0.95\columnwidth}{!}{\includegraphics{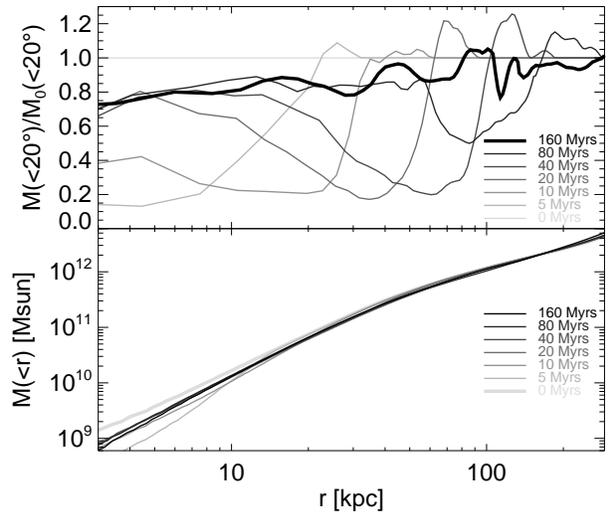}}
  \caption{{\em bottom panel:} cumulative mass $M_{\rm R}$ within
  radius $r$ as a function of $r$ at different times (grey scales);
  note the general inflation of the cluster atmosphere by the energy
  injection into the centre; {\em top panel:} ratio of target mass
  $M_{20}$ (mass contained within 20$^{\circ}$ of jet axis) over
  $M_{20}$ for the control simulation, indicating the relative gas
  depletion around the jet axis.  The residual $\sim 20\%$ reduction
  after 160 Myrs is consistent with the general reduction in density
  seen in the $M_{\rm R}$.\label{fig:massprofiles}}
\end{figure}

\begin{figure}
  \resizebox{0.95\columnwidth}{!}{\includegraphics{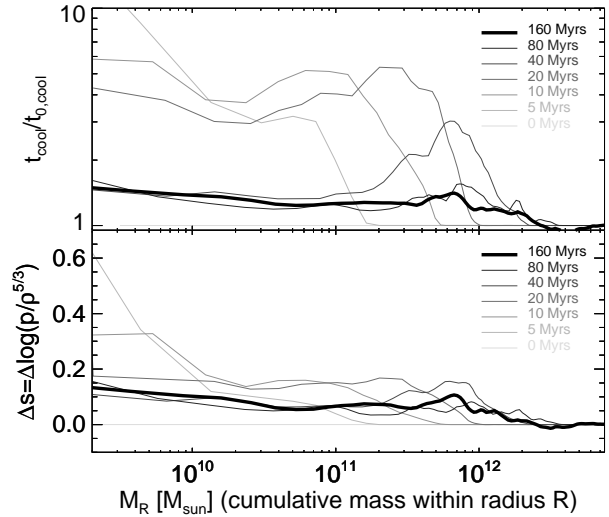}}
  \caption{{\em bottom panel:} average local increase in thermal
  cluster entropy $s=log(p/\rho^{5/3})$ relative to control
  simulation without jet, plotted as a function of cumulative mass
  $M(<r)$ (see Fig.~\ref{fig:massprofiles}).  {\em top panel:} ratio
  of average local cooling time of the thermal cluster gas with and
  without jet, showing the temporary suspension of
  cooling.\label{fig:lagrangian}}
\end{figure}

As found in earlier investigations of jet-cluster interactions, the
impact of the jet on the cluster leads to an increase in the central
entropy and a decrease in density, i.e, a net heat input.  This is
qualitatively apparent form the entropy panels in
Fig.~\ref{fig:maps1}.  Figures \ref{fig:massprofiles} and
\ref{fig:lagrangian} quantify this result.  The bottom panel of
Fig.~\ref{fig:massprofiles} shows the cumulative radial mass, i.e.,
the mass $M(<r)$ contained within radius $r$ from the cluster centre.
The initial time step lies above all other curves out to a radius of
about 60 kpc, indicating that the cluster has been inflated (i.e.,
mass moved out).

The bottom panel of Fig.~\ref{fig:lagrangian} shows the increase in
the mean entropy $\Delta s=\Delta \log{(p/\rho^{5/3})}$ of the cluster
gas at different time steps over the control simulation, plotted
against the Lagrangian coordinate $M(<r)$ from
Fig.~\ref{fig:massprofiles}.  This plot clearly shows that the entropy
of the central $10^{12}\,{\rm M_{\odot}}$ has been increased by about
0.1, corresponding to an increase in $p/\rho^{5/3}$ by about 25\%.

The top panel of Fig.~\ref{fig:lagrangian} shows the ratio of the
cooling times with and without jet.  At late times ($t \gtrsim
80\,{\rm Myrs}$) the cooling time for the inner few$\times
10^{12}\,{\rm M_{\odot}}$ has been increased by about 50\%, more at
earlier times.  This indicates that, on average, the jet can halt
cooling for a time significantly longer than the time the jet is
active.  Future simulations including radiative cooling will be
necessary to investigate whether this increase is sufficient to offset
cooling entirely and whether this result is consistent with {\em
XMM}-Newton and {\em Chandra} constraints on cluster cooling rates.

\begin{figure}
  \resizebox{0.95\columnwidth}{!}{\includegraphics{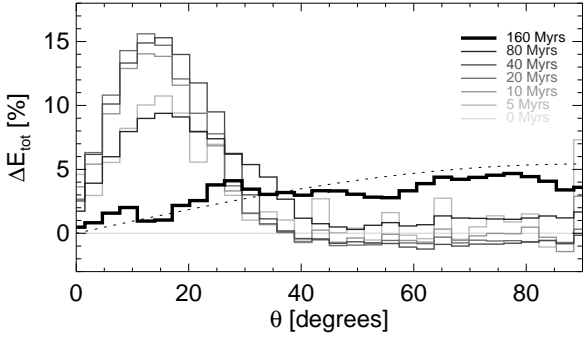}}
  \caption{Angular distribution (relative to mean jet axis) of excess
  energy $\Delta E$ injected by jet into thermal gas phase as
  percentage relative to total injected energy.  For comparison, the
  dotted line shows the isotropic case.\label{fig:angle}}
\end{figure}

\subsection{The isotropisation of energy and density}
One of the criticisms of models of cluster heating by jets is the
non-spherical nature of the energy input.  The problem is twofold:
Firstly, if the energy is predominantly injected into the axial
direction, the equatorial cluster gas does not benefit from the energy
injection and can cool unimpeded, while the axial gas is super-heated,
making AGN heating very inefficient.  Secondly, the channel of jet
exhaust and super-heated gas along the jet axis would allow subsequent
episodes of jet activity to punch easily through the inner cluster and
deposit their energy well outside the central region, thus allowing
the inner regions to cool unimpeded.

We already suggested that the shear and rotation in the central
cluster can rectify the second problem.  The top panel of
Fig.~\ref{fig:massprofiles} shows the radial profile of the target
mass $M_{20}$ contained in a $20^{\circ}$ angle from the jet axis,
relative to that of the control simulation.  Clearly, the amount of
mass is significantly reduced during the early stages of the
simulation.  However, at late times, $M_{20}$ increases back up to
80\% of the unperturbed value, consistent with the general reduction
of density in the central cluster shown in the bottom panel of
Fig.~\ref{fig:massprofiles}. Such a small reduction is dynamically
insignificant, i.e., the amount of target mass is more than sufficient
for a subsequent jet episode to couple to the cluster at {\em any}
radius.  This proves quantitatively that cluster dynamics can solve
the problem of evacuated channels raised by \cite{vernaleo:05}.

If cluster mass can be moved around to re-fill the jet channel and to
buffet the fossil radio plasma around, it is plausible to assume that
the energy input of the jet might be isotropised more rapidly as well.
The entropy panel of Fig.~\ref{fig:maps1} shows this qualitatively:
there is no clear high entropy channel along the axis of the jet.  In
fact, the draft of the rising bubbles draws low entropy material out
of the central cluster, as previously discussed by
\citep{churazov:03}.  This is borne out by a quantitative analysis:
Fig.~\ref{fig:angle} shows the excess energy in the thermal cluster
gas (once again excluding non-thermal jet exhaust) as a function of
polar angle $\theta$ away from the jet axis (measured around the
cluster centre).  At early times, the excess energy is clearly peaked
around the jet axis (within about 30$^{\circ}$),but at late times (160
Myrs, solid black line), the distribution is consistent with being
isotropic (dotted black line).  This indicates that individual
episodes of jet activity can, in fact, distribute their energy rather
isotropically into the cluster.  More detailed simulations including
radiative cooling and multiple jet episodes will be necessary to
develop a detailed picture of how heating and cooling in realistic
cluster atmospheres proceed.

\section{summary}
\label{sec:summary}
We have presented simulations of jet-cluster feedback that integrate a
correct representation of cluster dynamics, including dark matter, and
the collimated input of energy and momentum from an AGN.  The
simulations accurately reproduce the observed morphological appearance
of powerful radio sources like Cyg A, if a jitter is imposed on the
jet to account for unresolved dynamical instabilities of the jet
commonly referred to as the ``dentist drill effect''.  

We find that the dynamic structure of the cluster in the form of
rotation and shear in the velocity field and the inhomogeneity and
anisotropy of the cluster have significant impact on the evolution of
the jets and the radio lobes they inflate.  At late times, we find
that the excess energy deposited into the cluster by the action of the
jets is distributed well away from the axis, consistent with roughly
spherically symmetric energy input.  After long times (>160 Myrs), the
inner cluster has been sufficiently rearranged to essentially erase
the low density channel blasted out by the jet and move enough
material into the way of subsequent jet outbursts to couple
efficiently with the inner cluster.  This solves the problem of low
efficiency feedback found in simulations of spherically symmetric,
static atmospheres \citep{vernaleo:05}.

\thanks{We thank Volker Springel for providing us with a set of Gadget
simulated clusters.  We thank Mateusz Ruszkowski, Chris Reynolds,
Mitch Begelman, and Paul Nulsen for helpful discussions.  SH
acknowledges support by the National Aeronautics and Space
Administration through Chandra Postdoctoral Fellowship Award Number
PF3-40026 issued by the Chandra X-ray Observatory Center, which is
operated by the Smithsonian Astrophysical Observatory for and on
behalf of the National Aeronautics Space Administration under contract
NAS8-39073.  MB acknowledges support by DFG grant BR 2026/2
and the supercomputing grant NIC 1658 at the John von-Neumann centre
for computing at the Forschungszentrum J\"ulich. The software used in
this work was in part developed by the DOE-supported ASCI/Alliance
Center for Astrophysical Thermonuclear Flashes at the University of
Chicago.}


\begin{thebibliography}{23}
\expandafter\ifx\csname natexlab\endcsname\relax\def\natexlab#1{#1}\fi

\bibitem[{{Allen} {et~al.}(2006){Allen}, {Dunn}, {Fabian}, {Taylor}, \&
  {Reynolds}}]{allen:06}
{Allen}, S.~W., {Dunn}, R.~J.~H., {Fabian}, A.~C., {Taylor}, G.~B., \&
  {Reynolds}, C.~S. 2006, ArXiv Astrophysics e-prints

\bibitem[{{B{\^ i}rzan} {et~al.}(2004){B{\^ i}rzan}, {Rafferty}, {McNamara},
  {Wise}, \& {Nulsen}}]{birzan:04}
{B{\^ i}rzan}, L., {Rafferty}, D.~A., {McNamara}, B.~R., {Wise}, M.~W., \&
  {Nulsen}, P.~E.~J. 2004, \apj, 607, 800

\bibitem[{{Benson} {et~al.}(2003){Benson}, {Bower}, {Frenk}, {Lacey}, {Baugh},
  \& {Cole}}]{benson:03}
{Benson}, A.~J.~et al.~2003, \apj, 599, 38

\bibitem[{{Bower} {et~al.}(2006){Bower}, {Benson}, {Malbon}, {Helly}, {Frenk},
  {Baugh}, {Cole}, \& {Lacey}}]{bower:06}
{Bower}, R.~G.~et al.~2006, \mnras, 659

\bibitem[{{Br{\" u}ggen} \& {Kaiser}(2002{\natexlab{a}})}]{bruggen:02}
{Br{\" u}ggen}, M. \& {Kaiser}, C.~R. 2002{\natexlab{a}}, \nat, 418, 301

\bibitem[{{Br{\" u}ggen} \& {Kaiser}(2002{\natexlab{b}})}]{bruggen:02a}
---. 2002{\natexlab{b}}, \nat, 418, 301

\bibitem[{{Churazov} {et~al.}(2001){Churazov}, {Br{\" u}ggen}, {Kaiser}, {B{\"
  o}hringer}, \& {Forman}}]{churazov:01}
{Churazov}, E., {Br{\" u}ggen}, M., {Kaiser}, C.~R., {B{\" o}hringer}, H., \&
  {Forman}, W. 2001, \apj, 554, 261

\bibitem[{{Churazov} {et~al.}(2003){Churazov}, {Forman}, {Jones}, \& {B{\"
  o}hringer}}]{churazov:03}
{Churazov}, E., {Forman}, W., {Jones}, C., \& {B{\" o}hringer}, H. 2003, \apj,
  590, 225

\bibitem[{{Croton}(2005)}]{croton:05}
{Croton}, D.~J.~e.~a. 2005, \mnras, 356, 1155

\bibitem[{{Dalla Vecchia} {et~al.}(2004){Dalla Vecchia}, {Bower}, {Theuns},
  {Balogh}, {Mazzotta}, \& {Frenk}}]{dallavecchia:04}
{Dalla Vecchia}, C., {Bower}, R.~G., {Theuns}, T., {Balogh}, M.~L., {Mazzotta},
  P., \& {Frenk}, C.~S. 2004, \mnras, 507

\bibitem[{{Fabian} {et~al.}(2003){Fabian}, {Sanders}, {Crawford}, {Conselice},
  {Gallagher}, \& {Wyse}}]{fabian:03}
{Fabian}, A.~C., {Sanders}, J.~S., {Crawford}, C.~S., {Conselice}, C.~J.,
  {Gallagher}, J.~S., \& {Wyse}, R.~F.~G. 2003, \mnras, 344, L48

\bibitem[{{Forman} {et~al.}(2005){Forman}, {Nulsen}, {Heinz}, {Owen}, {Eilek},
  {Vikhlinin}, {Markevitch}, {Kraft}, {Churazov}, \& {Jones}}]{forman:05}
{Forman}, W.~et al.~2005, \apj, 635, 894

\bibitem[{{Fryxell} {et~al.}(2000){Fryxell}, {Olson}, {Ricker}, {Timmes},
  {Zingale}, {Lamb}, {MacNeice}, {Rosner}, {Truran}, \& {Tufo}}]{fryxell:00}
{Fryxell}, B.~et al.~2000, \apjs, 131, 273

\bibitem[{{Heinz} \& {Sunyaev}(2002)}]{heinz:02}
{Heinz}, S. \& {Sunyaev}, R. 2002, \aap, 390, 751

\bibitem[{{Lobanov} \& {Zensus}(2001)}]{lobanov:01}
{Lobanov}, A.~P. \& {Zensus}, J.~A. 2001, Science, 294, 128

\bibitem[{{Magorrian} {et~al.}(1998){Magorrian}, {Tremaine}, {Richstone},
  {Bender}, {Bower}, {Dressler}, {Faber}, {Gebhardt}, {Green}, {Grillmair},
  {Kormendy}, \& {Lauer}}]{magorrian:98}
{Magorrian}, J.~et al.~1998, \aj, 115, 2285

\bibitem[{{Nulsen} {et~al.}(2005){Nulsen}, {Hambrick}, {McNamara}, {Rafferty},
  {Birzan}, {Wise}, \& {David}}]{nulsen:05}
{Nulsen}, P.~E.~J.~et al.~2005, \apjl, 625, L9

\bibitem[{{Omma} \& {Binney}(2004)}]{omma:04b}
{Omma}, H. \& {Binney}, J. 2004, \mnras, 350, L13

\bibitem[{{Omma} {et~al.}(2004){Omma}, {Binney}, {Bryan}, \& {Slyz}}]{omma:04}
{Omma}, H., {Binney}, J., {Bryan}, G., \& {Slyz}, A. 2004, \mnras, 348, 1105

\bibitem[{{Quilis} {et~al.}(2001){Quilis}, {Bower}, \& {Balogh}}]{quilis:01}
{Quilis}, V., {Bower}, R.~G., \& {Balogh}, M.~L. 2001, \mnras, 328, 1091

\bibitem[{{Reynolds} {et~al.}(2001){Reynolds}, {Heinz}, \&
  {Begelman}}]{reynolds:01}
{Reynolds}, C.~S., {Heinz}, S., \& {Begelman}, M.~C. 2001, \apjl, 549, L179,
  rHB

\bibitem[{{Ruszkowski} {et~al.}(2004){Ruszkowski}, {Br{\" u}ggen}, \&
  {Begelman}}]{ruszkowski:04}
{Ruszkowski}, M., {Br{\" u}ggen}, M., \& {Begelman}, M.~C. 2004, \apj, 611, 158

\bibitem[{{Scheuer}(1974)}]{scheuer:74}
{Scheuer}, P.~A.~G. 1974, \mnras, 166, 513

\bibitem[{{Springel} {et~al.}(2001){Springel}, {White}, \&
  {Hernquist}}]{springel:01}
{Springel}, V., {White}, M., \& {Hernquist}, L. 2001, \apj, 549, 681

\bibitem[{{Vernaleo} \& {Reynolds}(2005)}]{vernaleo:05}
{Vernaleo}, J.~C. \& {Reynolds}, C.~S. 2005, ArXiv Astrophysics e-prints

\end{thebibliography}
\end{document}